\documentclass[aps,pra,showpacs]{revtex4-1}
\usepackage[dvips]{graphicx}
\usepackage{amsmath,amssymb,amsfonts,latexsym,color,dcolumn}
\begin{document}
\date{}
\title{Radiation from an oscillating dipole layer facing a conducting
plane: resonances and Dynamical Casimir Effect}
\author{C\'esar D. Fosco$^{1,2}$}
\author{Fernando C. Lombardo$^3$}
\affiliation{$^1$ Centro At\'omico Bariloche,
Comisi\'on Nacional de Energ\'\i a At\'omica,
R8402AGP Bariloche, Argentina}
\affiliation{$^2$ Instituto Balseiro,
Universidad Nacional de Cuyo,
R8402AGP Bariloche, Argentina}
\affiliation{$^3$  Departamento de F\'\i sica {\it Juan Jos\'e
Giambiagi}, FCEyN UBA and IFIBA CONICET-UBA, Facultad de Ciencias Exactas y Naturales,
Ciudad Universitaria, Pabell\' on I, 1428 Buenos Aires, Argentina}
%====================================================================
\begin{abstract} 
We study the properties of the classical electromagnetic (EM) radiation
produced by two physically different yet closely related systems, which
may be regarded as classical analogues of the Dynamical Casimir Effect
(DCE). 
They correspond to two flat, infinite, parallel
planes, one of them static and imposing perfect conductor boundary
conditions, while the other performs a rigid oscillatory motion. 
The systems differ just in the electrical properties of the oscillating
plane: one of them is just a planar dipole layer (representing, for
instance, a small-width electret).  The other, instead, has a dipole layer
on the side which faces the static plane, but behaves as a conductor on the
other side: this can be used as a representation of a conductor endowed
with patch potentials (on the side which faces the conducting plane). 

We evaluate, in both cases, the dissipative flux of energy between the
system and its environment, showing that, at least for small mechanical
oscillation amplitudes, it can be written in terms of the dipole layer
autocorrelation function.  We show that there are resonances as a function
of the frequency of the mechanical oscillation. 
\end{abstract}

\pacs{11.10.-z; 41.60.-m; 42.65.Yj; 42.50.-p}
\maketitle

%%%%%%%%%%%%%%%%%%%%%%%%%%%%%%%%%%%%%%%%%%%%%%%%%%%%%%%%%%%%%%%%%%%%%%%%%%
%%%%%%%%%%%%%%%%%%%%%%%%%%%%%%%%%%%%%%%%%%%%%%%%%%%%%%%%%%%%%%%%%%%%%%%%%%
%%%%%%%%%%%%%%%%%%%%%%%%%%%%% Introduction %%%%%%%%%%%%%%%%%%%%%%%%%%%%%%%
%%%%%%%%%%%%%%%%%%%%%%%%%%%%%%%%%%%%%%%%%%%%%%%%%%%%%%%%%%%%%%%%%%%%%%%%%% 
%%%%%%%%%%%%%%%%%%%%%%%%%%%%%%%%%%%%%%%%%%%%%%%%%%%%%%%%%%%%%%%%%%%%%%%%%%
\section{Introduction}\label{sec:intro}

The Dynamical Casimir Effect (DCE)~\cite{reviews dce},  provides a startling
example of the role that quantum fluctuations can play in the coupling between
the mechanical motion of a (neutral) mirror and the quantum degrees of
freedom of the EM field. That coupling may result, under appropriate
circumstances, in the resonant production of photons out of the vacuum. A
weaker, non-resonant effect may exist, in principle, even for the
case of a single accelerated mirror, where it may be understood as a
consequence of the conformal anomaly~\cite{Davies:1976hi}. 

A closer look at the role of the quantum fluctuations of the vacuum EM
field shows that they induce non-trivial correlation functions between charge
and current fluctuations in the media that compose the mirrors. Indeed, the
charge and current distributions have to rearrange themselves following a
fluctuation, in such a way that perfect-conductor boundary conditions hold
true.  The back-reaction of this rearrangement means that the charge and
current distributions, albeit having zero expectation value, acquire
non-trivial correlation functions.  These are objects with
a purely quantum origin, and therefore vanish when $\hbar \to 0$.

Note, however, that some systems exhibit {\em classical\/} charge or
current density autocorrelations. It is, therefore, important, to evaluate
the possible existence, as a consequence of them, of (classical)
motion-induced radiation. Besides its intrinsic interest, this phenomenon
could be relevant from the experimental point of view, since the effect may
be superposed to the DCE.  In a
previous work~\cite{Fosco:2014jha}, an example of such a situation has been
considered: a classical system consisting of  an oscillating plane endowed
with a dipole layer distribution may produce classical radiation when the
dipole layer is  autocorrelated. The classical autocorrelation considered
there was that of patch potentials~\cite{patch,Behunin1} on the moving surface,
although it could also correspond, for instance, to a plane with a
permanent polarization, like an electret.

One can also see this phenomenon from a different point of view:
it may be applied to obtain a classical, or rather semi-classical
realisation of the DCE. Indeed, as shown in~\cite{Fosco:2014jha}, the
quantum DCE is recovered when using a specific autocorrelation function
for the dipole layer distribution.

The work we present here may be regarded as an extension of that study to
configurations where there is an extra, conducting plane. This leads, as we
shall see, to a classical radiation that can exhibit resonances.  

We consider the classical radiation produced by two closely related
systems: the first and simplest one,  dealt with in
Sect.~\ref{sec:dipoles}, consists of an oscillating dipole layer \cite{ellis, heras} which
faces a static, perfectly-conducting plane.  The oscillating plane can
correspond, for instance, to a thin electret plane.  After introducing the
physical description  in~\ref{ssec:thedips}, in~\ref{ssec:rad} we calculate
the spectral density of radiation, understanding the latter as the
(spatially) averaged energy flux through a surface far from the oscillating
dipole layer. The calculation is performed in an approximate scheme,
keeping the lowest non-trivial contribution in an expansion in powers of
the amplitude of the oscillatory motion (assumed to be  much smaller than
the average distance between the planes). Then, in~\ref{ssec:andip} we
study and interpret different limits and particular cases of the general
result. 

In~\cite{Fosco:2014jha}, the result for the radiation due to a single dipole
layer has been used, with minor changes, to obtain the radiated power due
to patch potentials \cite{patch, Smoluchowski1, Pollack}. Here, the presence of the extra plane renders the
treatment of the patch potential case more complex. Therefore, in
Sect.~\ref{sec:patches}, we deal with that case, namely, also an
oscillating plane, facing a static perfectly conducting plane, such that
the former in endowed with a dipole layer but only on the side facing the
static plane, while the opposite side is instead a perfect conductor.
Contrary to what happened in the case of a single moving plane, the last
condition requires the introduction of an infinite number of extra image
charges (and therefore currents).  
The physical description of the model is introduced
in~\ref{ssec:thepatches}.  Because of the perfect-conductor condition on
the `external' face of the moving plane, there is no radiation:
the EM energy must be contained in between the two planes.  Thus we
consider, in~\ref{ssec:energy}, a different observable: the time derivative
of the energy contained between the planes as a function of time (also
using an expansion in the amplitude of oscillation). We evaluate that in
terms of the power that has to be applied to the moving plate.

In Sect.~\ref{sec:conclu}, we present our conclusions.

\section{Oscillating dipole layer}\label{sec:dipoles}
\subsection{The system}\label{ssec:thedips}

The configuration that we consider here consists of a static, perfectly
conducting plane, parallel to which a planar dipole layer performs a rigid
oscillatory motion along the direction defined by the normal to both
planes.  The position of the moving plane may be defined, by a suitable choice of
coordinates, in terms of a single `distance' function $\psi(t)$;
namely, $x_3 = \psi(t)$, while the static plane will be assumed to
correspond to $x_3=0$. 
Here, $x_3$ is one of the three spatial Cartesian coordinates of each
point ${\mathbf x} \equiv (x_1,x_2,x_3)$. With the choice of coordinates
mentioned in the previous paragraph, the dipole layer density $D$ is a
function of just two coordinates: $x_1$, $x_2$. Namely, $D = D({\mathbf
x}_\parallel)$, where we have introduced the notation ${\mathbf
x_\parallel} \equiv (x_1,x_2)$. 
The motion is assumed to be non-relativistic, so that the charge and
current densities due to the moving dipole layer, $\rho_D({\mathbf x},t)$
and ${\mathbf j}_D({\mathbf x},t)$, respectively, are given by:
\begin{eqnarray}\label{eq:sources}
\rho_D({\mathbf x},t) &=& - D({\mathbf x}_\parallel) \, \delta'(x_3-\psi(t)) \nonumber\\
{\mathbf j}_D({\mathbf x},t) &=& - D({\mathbf x}_\parallel)
\, \delta'(x_3-\psi(t)) \, {\dot \psi}(t)\,  \mathbf{\hat e_3} \;,
\end{eqnarray}
where $\mathbf{\hat e_3}$ is the unit vector along the direction of motion.

Finally, we shall assume that the motion is oscillatory around an average
distance $a$, so that there is a length $l$ such that $|\psi(t) - a| \leq
l$, $\forall t$. We will obtain approximate expressions under the
assumption that $l << a$.

We are mostly, although not exclusively, interested in the radiation due to
a dipole layer with a density that can be regarded as a random variable, so
that the resulting radiation flux may have a rather cumbersome spatial
dependence on the details of the layer.  Because of this, one will be
unable to detect the fine spatial details of such radiation; rather, one
should calculate global, coarse grained quantities, where those details are
averaged out. 

\subsection{Evaluation of the radiated power}\label{ssec:rad}

$U_{\rm rad}(x_3)$, the (average) radiated energy per unit area through a
constant-$x_3$ plane becomes:
\begin{equation}
	U_{\rm rad}(x_3)\;=\; \lim_{L\to\infty} \Big[ \frac{1}{L^2}\int
		dt\int_{|x_{1,2}|< L/2} d^2{\mathbf x}_{\parallel} \,
	S_3({\mathbf x}_\parallel,x_3,t)\Big]
\end{equation}
where $L^2$ is the area of the spatial plane (we take a $L \to \infty$
limit at the end to find the spatial average)  and $S_3$ is the ${\mathbf
e}_3$ component of the Poynting vector ${\mathbf S}=\frac{c}{4\pi}{\mathbf
E}\times {\mathbf B}$ (CGS Gaussian units are used throughout).  Note that,
in order to have a measure of the {\em radiated\/} energy, we will be
interested in $x_3 > {\rm max} \{ \psi(t)
\}$, discarding near-field, convective contributions, which decay with $x_3$. 

We see that the third component of ${\mathbf S}$ can be written as follows:
\begin{equation}\label{eq:s3}
S_3 \;=\; \frac{c}{4\pi} \, \epsilon_{ij} \, E_i B_j \;,
\end{equation}
where the indices $i$, $j$, as we shall henceforth assume, run from $1$
to $2$. To proceed, we need to write the electric and magnetic fields above in
terms of the charge and current densities given in (\ref{eq:sources}), in the presence of the
perfect-conductor boundary condition. For the geometry we are considering,
the boundary condition can be straightforwardly imposed by using images.
Indeed, the boundary condition at $x_3 = 0$ is automatically satisfied if
we include an image dipole layer, and solve for the fields (at any $x_3 >
0$) corresponding to the charge and current densities:
\begin{align}\label{eq:isources}
	&\rho({\mathbf x},t) = - D({\mathbf x}_\parallel) \Big[ \delta'(x_3-\psi(t)) 
+ \delta'(x_3 +\psi(t)) \Big]
\nonumber\\
& {\mathbf j}({\mathbf x},t) =- D({\mathbf x}_\parallel)
\big[ \delta'(x_3-\psi(t))-\delta'(x_3+\psi(t))\big]  {\dot \psi}(t)
\mathbf{\hat e_3} \;.
\end{align}

The fields $E_i$ and $B_i$ due to the sources above are more
straightforwardly found in terms of the scalar and vector potentials,
$\phi$ and ${\mathbf A}$, respectively:
\begin{eqnarray}
{\mathbf E}&=& -\nabla\phi-\frac{1}{c}\frac{\partial}{\partial t}{\mathbf A}\nonumber\\
{\mathbf B}&=&\nabla\times{\mathbf A} \;.
\end{eqnarray}
Indeed, using the Lorentz gauge-fixing condition, the potentials can be
expressed as follows:
\begin{eqnarray}
\phi({\mathbf x},t) &=& \int d^3{\mathbf x'}dt' \, G({\mathbf x},t;{\mathbf x'}, t') 
\, \rho({\mathbf x'},t') \nonumber\\
{\mathbf A}({\mathbf x},t) &=& \frac{1}{c} \, \int d^3{\mathbf x}'dt' 
G({\mathbf x},t;{\mathbf x'}, t')
{\mathbf j}({\mathbf x'},t') \;,
\end{eqnarray}
where $G$ denotes the retarded Green's function for the wave equation, 
which satisfies:
\begin{equation}
(c^{-2} \partial_t^2 - \nabla_{\mathbf x}^2) G({\mathbf x},t;{\mathbf
x'}, t') \;=\; 4\pi \, \delta({\mathbf x}-{\mathbf x'}) \, \delta(t-t') \;.
\end{equation}
A more explicit expression may be obtained by introducing the Fourier
transform: 
\begin{equation}
G({\mathbf x},t;{\mathbf x'}, t') \,=\, \int \frac{d\omega}{2\pi}
\frac{d^3{\mathbf k}}{(2\pi)^3} \; e^{ -i \omega (t -t') + i {\mathbf k}
\cdot ({\mathbf x}-{\mathbf x'})} \, {\widetilde G}({\mathbf k}_\parallel,k_3,\omega)
\;,
\end{equation}
with
\begin{equation}
{\widetilde G}({\mathbf k}_\parallel,k_3,\omega)=\frac{4\pi}{{\mathbf
k}_\parallel^2+k_3^2-(\frac{\omega}{c}+i\eta)^2} \;\,
\end{equation}
and $\eta$ denotes and infinitesimal positive constant. 

Since ${\mathbf A}$ points to the $ \mathbf{\hat e_3}$ direction, we see
that the components of the electric and magnetic field relevant to
the calculation of (\ref{eq:s3}) are given by:
\begin{equation}
E_i \,=\, - \partial_i \phi \;,\;\; B_i \,=\, \epsilon_{ij} \partial_j A_3
\;,
\end{equation}
so that 
\begin{equation}
S_3 \,=\,\frac{c}{4\pi} \, \partial_j \phi \, \partial_j A_3 \;. 
\end{equation} 

Then, we get for $U_{\rm rad}(x_3)$ the expression:
\begin{align}\label{eq:ur1}
	 U_{\rm rad}(x_3)&=\, \lim_{L\to\infty} \Big\{ \frac{1}{4 \pi L^2}
	\int dt\int_{|x_{1,2}|< L/2} d^2{\mathbf x}_{\parallel} \int dt'
	\int d^3{\mathbf x}
	\int dt'' \int d^3{\mathbf x''} \Big[ 
\frac{\partial G}{\partial x_j}({\mathbf x},t;{\mathbf x'}, t') 
 \frac{\partial G}{\partial x_j}({\mathbf x},t;{\mathbf x''}, t'')
 \nonumber\\
 & \times D({\mathbf x_\parallel'})  D({\mathbf x_\parallel''}) \big(
 \delta'(x'_3-\psi(t')) + \delta'(x'_3 +\psi(t')) \big) \big(
 \delta'(x''_3-\psi(t''))-\delta'(x''_3+\psi(t''))\big)
  \dot{q}(t'') \Big]\Big\} \;.
\end{align}

We then perform the integrals over $t$, ${\mathbf x_\parallel}$,
${\mathbf x'_\parallel}$ and ${\mathbf x''_\parallel}$, obtaining a result
that may be conveniently written as follows:
\begin{equation}\label{eq:ur2}
U_{\rm rad}(x_3)\,=\,\frac{1}{4 \pi} \int \frac{dk_3}{2\pi}  \int
\frac{dp_3}{2\pi}\, e^{i x_3 (k_3 + p_3)} \, 
\int\frac{d\omega}{2\pi}\, \int \frac{d^2{\mathbf k_\parallel}}{(2\pi)^2}
\,\Big[ {\mathbf k_\parallel}^2
 \widetilde{\Omega}({\mathbf k_\parallel}) 
\, \widetilde{G}(\omega,{\mathbf k_\parallel},k_3) 
\, \widetilde{G}(-\omega,-{\mathbf k_\parallel},p_3) 
 \, \Lambda(\omega,k_3,p_3) \Big]
\end{equation}
where
\begin{align}
 \Lambda(\omega,k_3,p_3)\,=\,\int dt' \int dt''  \int dx'_3 \int dx''_3 \,
 e^{i[\omega(t'- t'')- (k_3 x'_3 + p_3 x''_3)]} \, 
& \big(\delta'(x'_3-\psi(t')) + \delta'(x'_3 +\psi(t')) \big) 
\nonumber\\
\times \big( \delta'(x''_3-\psi(t''))-\delta'(x''_3+\psi(t''))\big)
 &\dot{\psi}(t'') \;,
 \end{align}
and we have introduced $\widetilde{\Omega}({\mathbf k_\parallel})$, the
Fourier transform of the autocorrelation function for the dipole layer:
\begin{equation}
\Omega({\mathbf x}_\parallel)\,=\, \frac{1}{L^2} \, 
\int d^2 y_\parallel  D({\mathbf y}_\parallel) D({\mathbf
x}_\parallel+{\mathbf y}_\parallel)\;.
\end{equation}

In natural ($\hbar =1$ and $c=1$) units, ${\widetilde \Omega}$ above is a
dimensionless quantity.  Note that a similar expression to the one above
could have been obtained if one had a random patch potential distribution,
with a translation invariant stochastic correlation \cite{patch}. Namely, even without
evaluating the average over a constant-$x_3$ plane, the translation
invariance of the system does produce an entirely analogous expression to
the one above, now interpreting $\Omega$ as the result of an average with a
statistical weight.

We next evaluate $\Lambda(\omega,k_3,p_3)$ to the second order in $q(t)$,
the departure of $\psi(t)$ from its average position $a$;
namely, $\psi(t)=a + q(t)$:
\begin{equation}
	\Lambda(\omega,k_3,p_3) \,\sim\,\Lambda^{(2)}(\omega,k_3,p_3) 
\end{equation}
with
\begin{equation}
	\Lambda^{(2)}(\omega,k_3,p_3)\,=\,
	- |\tilde{q}(\omega)|^2 \omega \, k_3^2 \, p_3 \,
	(e^{i k_3 a} - e^{-i k_3 a})  
	\, (e^{i p_3 a} - e^{-i p_3 a}) \;.  
\end{equation}
This order is the first non-trivial one to produce a non-vanishing
contribution to the radiated energy.

We then evaluate the integrals over $k_3$ and $p_3$, which can be
performed, for example, by using Cauchy's theorem in a rather
straightforward way, obtaining a result that, contains both convection and
radiation terms.

Keeping just the radiation terms, we find:
\begin{equation}\label{eq:desp0}
U_{\rm rad} \,=\, \int_0^\infty \frac{d\omega}{2\pi} \, {\mathcal P}({\omega}) \;,
\end{equation}
where the spectral density ${\mathcal P}({\omega})$ is given by:
\begin{equation}\label{eq:desp1}
{\mathcal P}(\omega)\, = \,  8 \pi \vert\omega\vert
\vert\tilde{q}(\omega)\vert^2 \,
\int \frac{d^2{\mathbf k_\parallel}}{(2\pi)^2} \theta\left(\frac{\vert \omega
\vert}{c} - \vert {\mathbf k_\parallel}\vert\right) \, {\mathbf k_\parallel}^2 \; 
\sqrt{(\frac{\omega}{c})^2-{\mathbf k_\parallel}^2}
\;
\sin^2[ a \sqrt{(\frac{\omega}{c})^2-{\mathbf k_\parallel}^2}]
\;
\widetilde{\Omega}({\mathbf k_\parallel}) \,.
\end{equation}
And, assuming the autocorrelation function to be isotropic,
\begin{equation}\label{eq:desp}
{\mathcal P}(\omega)\, = \,  4 \vert\omega\vert\vert\tilde
q(\omega)\vert^2\int_0^{\omega/c}dk_\parallel\,
k_\parallel^3 \; 
\sqrt{(\frac{\omega}{c})^2-k_\parallel^2}
\;
\sin^2[ a \sqrt{(\frac{\omega}{c})^2-k_\parallel^2}]
\;
\widetilde{\Omega}(k_\parallel) \;, 
\end{equation}
which is the main result of this Section. 

As an example, we consider a sharp-cutoff model for the autocorrelation
function ${\tilde \Omega}(k_\parallel)$ given by \cite{patch}
\begin{equation}
{\tilde \Omega}(k_\parallel) = \frac{4 d^2}{k_{\rm max}^2 - k_{\rm min}^2}
\theta(k_{\rm max} - k_\parallel) \theta(k_\parallel - k_{\rm min})\;,\label{sharp}
\end{equation} 
identical to the one used in~\cite{patch} within the context of patch
potentials but here interpreted in the context of polarization correlation
function.
In this case, the spectral density in Eq.(\ref{eq:desp}) is 

\begin{eqnarray}
{\mathcal P}(\omega)\, & = & \, \frac{2}{15} \frac{ d^2 \vert \omega\vert \vert {\tilde q}(\omega)\vert^2}{(k_{\rm max}^2 - k_{\rm min}^2)} \left[ \frac{4\omega^2}{c^2} \left(k_{\rm min}^2 \sqrt{(\frac{\omega}{c})^2 - k_{\rm min}^2} - k_{\rm max}^2 \sqrt{(\frac{\omega}{c})^2 - k_{\rm max}^2}\right) \right. \nonumber \\
&+& 12 \left(k_{\rm max}^4 \sqrt{(\frac{\omega}{c})^2 - k_{\rm max}^2} - k_{\rm min}^4 \sqrt{(\frac{\omega}{c})^2 - k_{\min}^2}\right) \nonumber \\
& + & 8 \frac{\omega^4}{c^4} \left(\sqrt{(\frac{\omega}{c})^2 - k_{\rm min}^2} - \sqrt{(\frac{\omega}{c})^2 - k_{\max}^2}\right) \nonumber \\
& + & \frac{30}{a^4} \left(\frac{\omega^2a^2}{c^2} - (3 + 2 a^2k_{\rm min}^2)\right) \sqrt{(\frac{\omega}{c})^2 - k_{\rm min}^2} \cos\left[2a \sqrt{(\frac{\omega}{c})^2 - k_{\rm min}^2}\right] \nonumber \\
&- & \frac{30}{a^4} \left(\frac{\omega^2a^2}{c^2} - (3 + 2 a^2k_{\rm max}^2)\right) \sqrt{(\frac{\omega}{c})^2 - k_{\rm max}^2} \cos\left[2a \sqrt{(\frac{\omega}{c})^2 - k_{\rm max}^2}\right] \nonumber \\
&+&  \frac{30}{a^3}  \left(\frac{\omega^2a^2}{c^2} - 3\right) \left(k_{\rm max}^2  \sin\left[2a \sqrt{(\frac{\omega}{c})^2 - k_{\rm max}^2}\right] - k_{\rm min}^2  \sin\left[2a \sqrt{(\frac{\omega}{c})^2 - k_{\rm min}^2}\right] \right) \nonumber \\
&+& \frac{15}{a^5}  \left(5 \frac{\omega^2a^2}{c^2} - 3\right) \left( \sin\left[2a \sqrt{(\frac{\omega}{c})^2 - k_{\rm max}^2}\right] -  \sin\left[2a \sqrt{(\frac{\omega}{c})^2 - k_{\rm min}^2}\right] \right) \nonumber \\
&+& \left. \frac{30}{a}  \left(k_{\rm min}^4  \sin\left[2a \sqrt{(\frac{\omega}{c})^2 - k_{\rm min}^2}\right] - k_{\rm max}^4  \sin\left[2a \sqrt{(\frac{\omega}{c})^2 - k_{\rm max}^2}\right] \right) 
\right] 
\end{eqnarray}
where it has been assumed that $k_{\rm min} < \omega/c < k_{\rm max}$, for arbitrary cutoff-scales.

\subsection{Analysis of the result}\label{ssec:andip}

Let us consider here some particular cases of the general
result: The first  amounts to the situation (that we had already
considered in~\cite{Fosco:2014jha}) of an autocorrelation function
$\widetilde{\Omega}$ such that the corresponding correlation length is much
smaller than $c/\omega$. Then, $\widetilde{\Omega}({\mathbf
k}_\parallel)$ can be approximated by $\widetilde{\Omega}({\mathbf 0})$,
and extracted out of the integral. Thus,
\begin{eqnarray}\label{eq:shortcorr}
{\mathcal P}(\omega) & \simeq &   4 \,
\widetilde{\Omega}({\mathbf 0}) \; \vert\omega\vert\vert\tilde
q(\omega)\vert^2\int_0^{\omega/c}dk_\parallel\, k_\parallel^3 \;
\sqrt{(\frac{\omega}{c})^2-k_\parallel^2} \; \sin^2[ a
\sqrt{(\frac{\omega}{c})^2-k_\parallel^2}]\nonumber\\
& \simeq &   \frac{4}{15 c^5}\, p^2 \;|\omega|^6 \; |\tilde{q}(\omega)|^2
\nonumber\\
&+& \frac{c}{a^6}\, p^2 \;\Big\{ 
(\frac{\omega a}{c})^2 \big[(\frac{\omega a}{c})^2 - 3\big] \; \cos(\frac{2\omega a}{c})
\,-\, \frac{1}{2} \, |\frac{\omega a}{c}| \big[ 5 (\frac{\omega a}{c})^2 -
3] \; \sin(|\frac{2\omega a}{c}|)\big]
\Big\} \; |\tilde{q}(\omega)|^2
\end{eqnarray}
where we have used the notation $\widetilde{\Omega}({\mathbf 0}) = p^2$,
since it has the dimensions and interpretation of a single dipole moment $\vec p$.
Note that the first, $a$-independent contribution in the previous
expression for the power, is twice the value we had found in the absence of
the conducting plane. This is understood as follows: here we have two
dipoles, the real one and its image, which yields a factor of four. But in
our previous reference we had radiation both for positives and negative
values of $x_3$, while here the conductor at $x_3=0$ means that only the
positive $x_3$ contribution has to be dealt with. Thus there must be a factor of
$ 4/2 = 2$ between the first line of (\ref{eq:shortcorr}) and the analogous result
in the absence of the conductor, which is correct.

The next term in (\ref{eq:shortcorr}) has a richer structure, since it
introduces resonant peaks in the total power, the location of which are
the roots of a transcendent equation. It is convenient to introduce
${\mathcal P}_a(\omega)$, the part of the power which depends on $a$:
\begin{equation}\label{eq:shortcorr1}
{\mathcal P}_a(\omega) \; \simeq \; \frac{c}{a^6} p^2 \, f(\xi) \, |\tilde{q}(\omega)|^2
\end{equation}
where $\xi \equiv \frac{\omega a}{c}$ and
\begin{equation}\label{eq:deff}
f(\xi) \;=\;  \xi^2 (\xi^2 - 3) \; \cos(2 \xi) \,-\, \frac{1}{2} \,
|\xi| (5 \xi^2 - 3) \; \sin(2|\xi|) 
\;.
\end{equation}
In Fig. 1 we plot the function $f(\xi)$: we see that the 
interference between the radiation emitted by the dipole layer
and its reflection on the static mirror implies the existence of peaks in
the radiated power. 
 
 \begin{figure}[!ht]
\includegraphics[width=12cm]{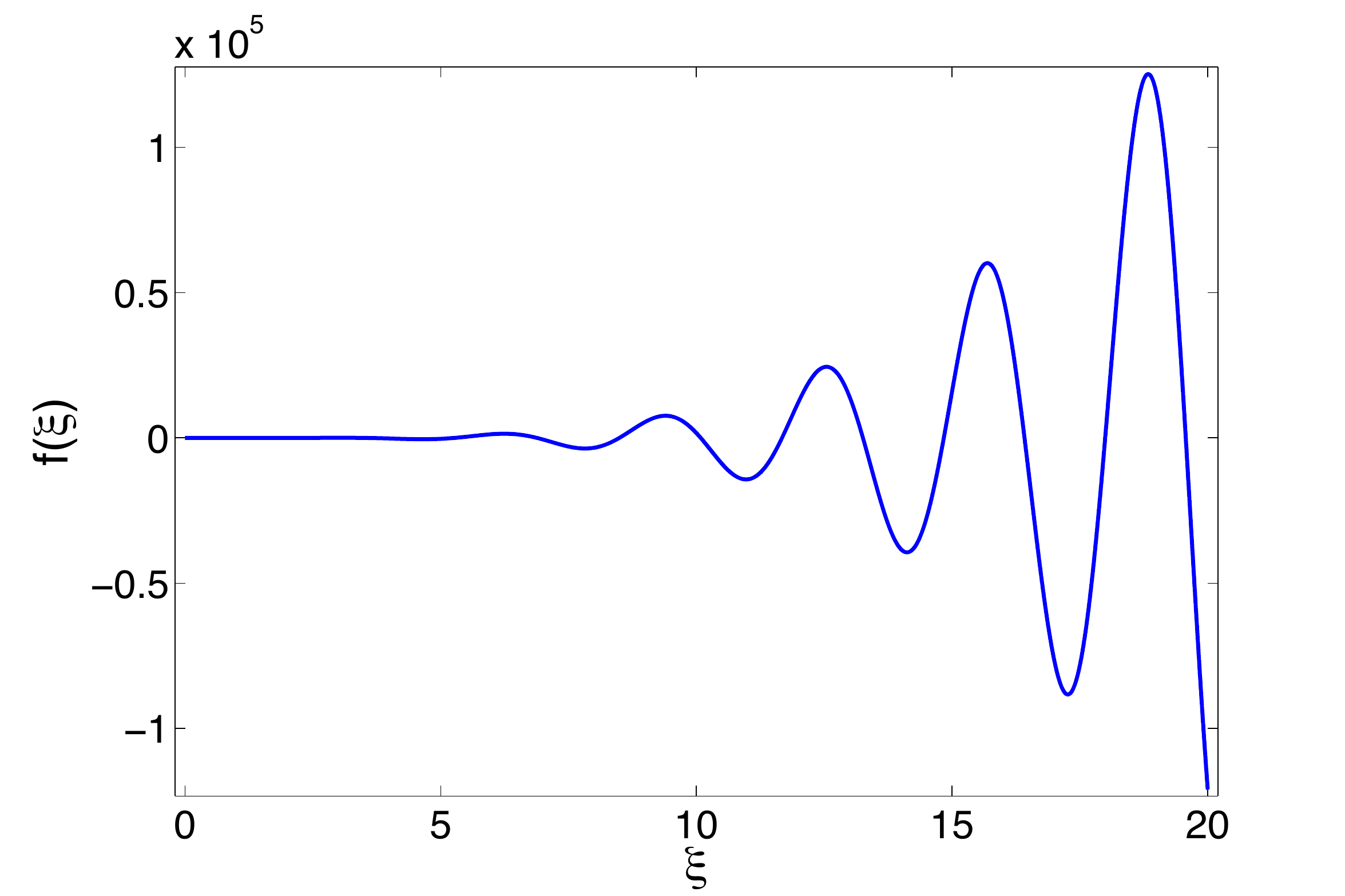}
\caption{(Color online) The function $f(\xi)$ ($\xi \equiv \frac{\omega a}{c}$),
defined in Eq.~(\ref{eq:deff}), which determines the $a$-dependent part of
the radiated power per unit area for a given mechanical frequency $\omega$.}
\label{fig1}
\end{figure}

The second particular case corresponds to a more immediate limit of the
general result, and corresponds to $a \to \infty$ (we recall that the power
is always evaluated at $x_3 >> a$) and the result becomes
independent of $x_3$). In this situation, we have:
\begin{equation}\label{eq:longa}
{\mathcal P}(\omega)\,\sim\, 2 \vert\omega\vert\vert\tilde
q(\omega)\vert^2\int_0^{\omega/c}dk_\parallel\,
k_\parallel^3 \; 
\sqrt{(\frac{\omega}{c})^2-k_\parallel^2} \;
\widetilde{\Omega}(k_\parallel) \;,
\end{equation}
since the highly oscillating sine function (assuming $\widetilde{\Omega}$
to be a smooth function of its argument) can be approximated by $1/2$. This
corresponds again to twice the result in the absence of the conducting
plane, and for the same reason. It is also self-evident that all the
resonant phenomena disappear in this limit. 

Using the sharp-cutoff model for the autocorrelation function ${\tilde \Omega}(k_\parallel)$ (with $k_{\rm max} > \omega/c$) 
it is possible to find that the power in (\ref{eq:longa}) is 
${\mathcal P}(\omega) \sim \vert\tilde q(\omega)\vert^2 \omega^6/c^5$. This case, can be compared with that coming 
from the DCE \cite{FLM2013}, since a single accelerated perfect mirror creates photons due to the interaction with 
the quantum fluctuations of the electromagnetic field. Just considering dimensional arguments, and limiting the analysis to the 
non-relativistic case,  one expect that the dissipative force per unit length on the mirror to be proportional to $\dddot a/c^4$. 
This force corresponds to a spectral density $ {\mathcal P}_{\rm DCE}(\omega) 
\sim\vert\tilde q(\omega)\vert^2 \omega^6/c^4$ \cite{reviews dce}.  

Finally, as the final limiting case we see that the radiated energy also
vanishes when the correlation length is infinite. This result can be
understood by applying Gauss law for ${\mathbf E}$, on a large closed box with
two of its faces parallel to the planes, taking into account the fact that
the total electric charge is zero.
%%%%%%%%%%%%%%%%%%%%%%%%%%%%%%%%%%%%%%%%%%%%%%%%%%%%%%%%%%%%%%%%%%%%%%%%%%
%%%%%%%%%%%%%%%%%%%%%%%%%%%%%%%%%%%%%%%%%%%%%%%%%%%%%%%%%%%%%%%%%%%%%%%%%%
%%%%%%%%%%%%%%%%%%%%%%%%%%% Patch potentials %%%%%%%%%%%%%%%%%%%%%%%%%%%%%
%%%%%%%%%%%%%%%%%%%%%%%%%%%%%%%%%%%%%%%%%%%%%%%%%%%%%%%%%%%%%%%%%%%%%%%%%%
%%%%%%%%%%%%%%%%%%%%%%%%%%%%%%%%%%%%%%%%%%%%%%%%%%%%%%%%%%%%%%%%%%%%%%%%%%
\section{Patch potentials}\label{sec:patches}
\subsection{The system}\label{ssec:thepatches}

Let us consider the second model, again assuming the motion to be
non-relativistic. The moving plane is assumed to have a patch potential
distribution on its side facing the static mirror. 

Now, the patch potentials may be constructed by introducing two dipole
layers close to the moving plane. One of them represents the patch potential itself, and is
located at $x_3 = \psi^-(t)$,  $\psi^- \equiv \psi - \epsilon$ ($\epsilon$ a
positive infinitesimal). The other, its image,  is introduced in order
to have perfect conductor boundary conditions at $x_3 = \psi(t)$;
therefore it is located at $\psi^+ \equiv \psi+ \epsilon$. Leaving aside for the moment
the existence of perfect boundary conditions at $x_3 = 0$, we have the 
charge and current densities $\rho_P({\mathbf x},t)$
and ${\mathbf j}_P({\mathbf x},t)$, respectively:
\begin{eqnarray}\label{eq:psources}
\rho_P({\mathbf x},t) &=& - D({\mathbf x}_\parallel) \, \big[
\delta'(x_3-\psi^-(t)) + \delta'(x_3-\psi^+(t)) \big] \;\,,\nonumber\\
{\mathbf j}_P({\mathbf x},t) &=& - D({\mathbf x}_\parallel) \, \big[
\delta'(x_3-\psi^-(t)) + \delta'(x_3-\psi^+(t)) \big]
\, {\dot \psi}(t)\,  \mathbf{\hat e_3} \;.
\end{eqnarray}
In order to satisfy perfect conductor boundary conditions at $x_3=0$, we
must introduce now an infinite number of images which mirror the previous
sources (at the appropriate locations). Is is straightforward to see that
the full charge and current densities which ensure the proper boundary
conditions to hold true, are given by:
\begin{eqnarray}\label{eq:pesources}
\rho({\mathbf x},t) &=& - D({\mathbf x}_\parallel) \,
\sum_{n=-\infty}^{+\infty} 
\big[ \delta'(x_3-\psi_n^-(t)) + \delta'(x_3-\psi_n^+(t)) \big] \;\,,\nonumber\\
{\mathbf j}({\mathbf x},t) &=& - D({\mathbf x}_\parallel) \,
\sum_{n=-\infty}^{+\infty} \big[ \delta'(x_3-\psi_n^-(t)) + \delta'(x_3-\psi_n^+(t)) \big] \,
\, {\dot \psi}_n(t)\,  \mathbf{\hat e_3} \;,
\end{eqnarray}
where: $\psi_n(t) \equiv (2 n + 1) \psi(t)$, and $\psi_n^\pm(t) \equiv (2 n
+ 1) \psi^\pm (t)$ . 

It is convenient, for later use, to note that the charge and current
densities above may be written as follows:
\begin{equation}\label{eq:altsources}
\rho({\mathbf x},t) \,=\, - D({\mathbf x}_\parallel) \,\frac{\partial
\sigma}{\partial x_3} (x_3,t) \;\;,\;
{\mathbf j}({\mathbf x},t) \,=\,  D({\mathbf x}_\parallel) \,
\frac{\partial \sigma}{\partial t}(x_3,t) \, \mathbf{\hat e_3} \;,
\end{equation}
with:
\begin{equation}\label{eq:defsigma}
\sigma(x_3,t) \;=\; \sum_{n=-\infty}^{+\infty} \big[
\delta(x_3-\psi_n^-(t)) + \delta(x_3-\psi_n^+(t)) \big] \;. 
\end{equation}

\subsection{Mechanical power per unit area on the moving
plate}\label{ssec:energy}

There is no radiation outside of the volume enclosed by the two
plates, but there may be energy traded between the EM field and the
environment,  by means of the mechanical work exerted on the moving mirror.
The rate of change of ${\mathcal E}(t)$, the total energy per unit area contained
between the static and moving planes, is given by:
\begin{equation}\label{eq:dedt}
\frac{d{\mathcal E}}{dt}(t) \;=\; \dot{\psi}(t) \, f(t)
\end{equation}
where $f(t)$ denotes the $x_3$ component of the force per unit area on the moving 
plate. This is in turn obtained as follows:
\begin{equation}
f(t) \;=\; \int d^2{\mathbf x}_\parallel \, \big[
T_{33}({\mathbf x}_\parallel, \psi^+(t),t) \,-\, T_{33}({\mathbf
x}_\parallel,\psi^-(t),t) \big] 
\end{equation}
where $T_{ij}$ is Maxwell's stress tensor. 

In terms of the potentials, $T_{33}$, the only relevant
component of $T_{ij}$ for this calculation, may be written as follows:
\begin{equation}
T_{33} \;=\; - \frac{1}{8\pi} \,\Big[(\partial_j\partial_3 \chi)^2 \,+\,  
\frac{1}{c^2} (\partial_j \partial_t \chi)^2 \,-\,
(\partial_3^2\chi + \frac{1}{c^2}\partial_t^2 \chi)^2 \Big] \;.  
\end{equation}
where we have used the property that the potentials may be written (for the kind of
sources that we are considering) as follows:
\begin{equation}
\phi  \;=\;- \frac{\partial\chi}{\partial x_3}
\;\;,\;\;\;\;
A_3 \;=\;  \frac{1}{c} \frac{\partial\chi}{\partial t} \;,
\end{equation}
where the scalar function $\chi$ can be written in terms of the retarded
Green's function $G$:
\begin{equation}
\chi({\mathbf x},t)\;=\; \int d^3{\mathbf x'} \int dt' \;
G({\mathbf x},t;{\mathbf x'},t') \,D({\mathbf x'_\parallel})
\sigma(x'_3,t') \;,
\end{equation}
with $\sigma(x_3,t)$ as introduced previously in (\ref{eq:defsigma}). 

The perturbative expansion for $\frac{d{\mathcal E}(t)}{dt} = \dot{\mathcal
E}(t)$, in powers of $q(t)$ has the form
\begin{equation}
\dot{\mathcal E}(t) \;=\; \dot{\mathcal E}^{(0)}(t) \,+\, \dot{\mathcal E}^{(1)}(t)
\,+\, \dot{\mathcal E}^{(2)}(t) \,+\,\ldots 
\end{equation}
where the index denotes the order of the corresponding term. Since
$\dot{\psi}$ is of order $1$, we see that:
\begin{equation}
\dot{\mathcal E}^{(0)}\;=\;0 \;\;,\;\;
\end{equation}
as it should be, since at the zeroth order both planes are static, and
therefore there can be no power. 

Regarding the first-order term, one clearly has the relation:
\begin{equation}
\dot{\mathcal E}^{(1)} \;=\; \dot{\psi}(t) \,f^{(0)} \;,  
\end{equation}
where $f^{(0)}$ denotes the zeroth order force per unit area for a static
system. Thus, the previous equation may be integrated out to obtain:
\begin{equation}
\delta{\mathcal E}\;=\; \delta a \, f^{(0)} \;,  
\end{equation}
in other words, $f^{(0)} = (\frac{\partial {\mathcal E}}{\partial a})^{(0)}$. 
The explicit evaluation of $f^{(0)}$ yields:
\begin{equation}
f^{(0)}\;=\; 8 \pi \, \int \frac{d^2{\mathbf k}_\parallel}{(2\pi)^2}
\, |{\mathbf k}_\parallel|^2 \, {\widetilde{\Omega}}({\mathbf k}_\parallel)
\, \Big[ \frac{1}{e^{ 2 |{\mathbf k}_\parallel| a}-1} + 
\frac{1}{(e^{ 2 |{\mathbf k}_\parallel| a}-1)^2} \Big] \;,
\end{equation} 
which agrees with the force one would obtain from the static interaction energy 
resulting between two conductors with patch potentials~\cite{FLM2013}.

The second order term, requires the calculation of $f^{(1)}$:
\begin{equation}
\dot{\mathcal E}^{(2)} \;=\; \dot{\psi}(t) \,f^{(1)}(t) \;.  
\end{equation}
We see that:
\begin{align}
f^{(1)}(t) \;=\;\frac{1}{4\pi}\, \lim_{L\to\infty} \frac{1}{L^2} \,
\int_{-L/2}^{+L/2} dx_1 \int_{-L/2}^{+L/2} dx_2 \,\Big\{
& - \big[\partial_j\partial_3\chi(x_1,x_2,\psi^-(t),t)\big]^{(0)} 
    \big[\partial_j\partial_3\chi(x_1,x_2,\psi^-(t),t)\big]^{(1)} \\ 
& - \big[\frac{1}{c} \partial_j\partial_t\chi(x_1,x_2,\psi^-(t),t)\big]^{(0)} 
 \big[\frac{1}{c} \partial_j\partial_t\chi(x_1,x_2,\psi^-(t),t)\big]^{(1)} \\ 
&+ \big[ \partial_3^2\chi(x_1,x_2,\psi^-(t),t) - \frac{1}{c^2}\partial_t^2
\chi(x_1,x_2,\psi^-(t),t) \big]^{(0)} \\
& \times \big[ \partial_3^2\chi(x_1,x_2,\psi^-(t),t) - \frac{1}{c^2}\partial_t^2
\chi(x_1,x_2,\psi^-(t),t) \big]^{(1)} \Big\} \;.
\end{align}

It is rather straightforward to show that the first two terms on the rhs of
the equation above vanish. Thus, keeping only the third term, and
introducing Fourier transforms, we see that:
\begin{equation}
f^{(1)}(t) = \frac{1}{4\pi} \, \int \frac{d^2{\mathbf
k}_\parallel}{(2\pi)^2}\, {\widetilde{\Omega}}({\mathbf k}_\parallel)
\, F_1({\mathbf k}_\parallel,a) \; 
F_2 ({\mathbf k}_\parallel,a,t) \;,
\end{equation}
where we have introduced:
\begin{equation}
F_1({\mathbf k}_\parallel,a) \;\equiv\; \int \frac{dk_3}{2\pi}
\int \frac{d\omega}{2\pi} \, e^{i ( k_3 a^- - \omega t)} 
\, \widetilde{G}({\mathbf k}_\parallel,k_3,0) \,k_3^2 \,
\widetilde{\sigma}^{(0)}(k_3,\omega) \;,  
\end{equation}
and
\begin{equation}
F_2({\mathbf k}_\parallel,a,t) \;\equiv\; \int \frac{dk_3}{2\pi}
\int \frac{d\omega}{2\pi} \, e^{i ( k_3 a^- - \omega t)} 
\, \widetilde{G}({\mathbf k}_\parallel,k_3,\omega) \,[k_3^2
-(\frac{\omega}{c})^2]\,
\widetilde{\sigma}^{(1)}(k_3, \omega) \;.  
\end{equation}
Using the explicit form of the Green's functions,  $\sigma$, and of their
Fourier transforms, we find after some algebra:
\begin{equation}
F_1({\mathbf k}_\parallel,a) \;=\; - 4 \pi \; |{\mathbf k}_\parallel| \; \coth(|{\mathbf
k}_\parallel| a) \;,
\end{equation}
while for $F_2$ the result may be put in the form:
\begin{equation}
F_2({\mathbf k}_\parallel,a,t) = 
 \frac{4 \pi}{a^2} \,|{\mathbf k}_\parallel|^2\, 
\,(-1 + a \frac{d}{da})
\sum_{-\infty}^{+\infty} \int \frac{d\omega}{2\pi} \,
\, e^{-i \omega t} \frac{\tilde{q}(\omega)}{(\frac{n \pi}{a})^2 + {\mathbf
k}_\parallel^2 - (\frac{\omega + i
\eta}{c})^2} \;. 
\end{equation}

 \begin{figure}[!ht]
\includegraphics[width=12cm]{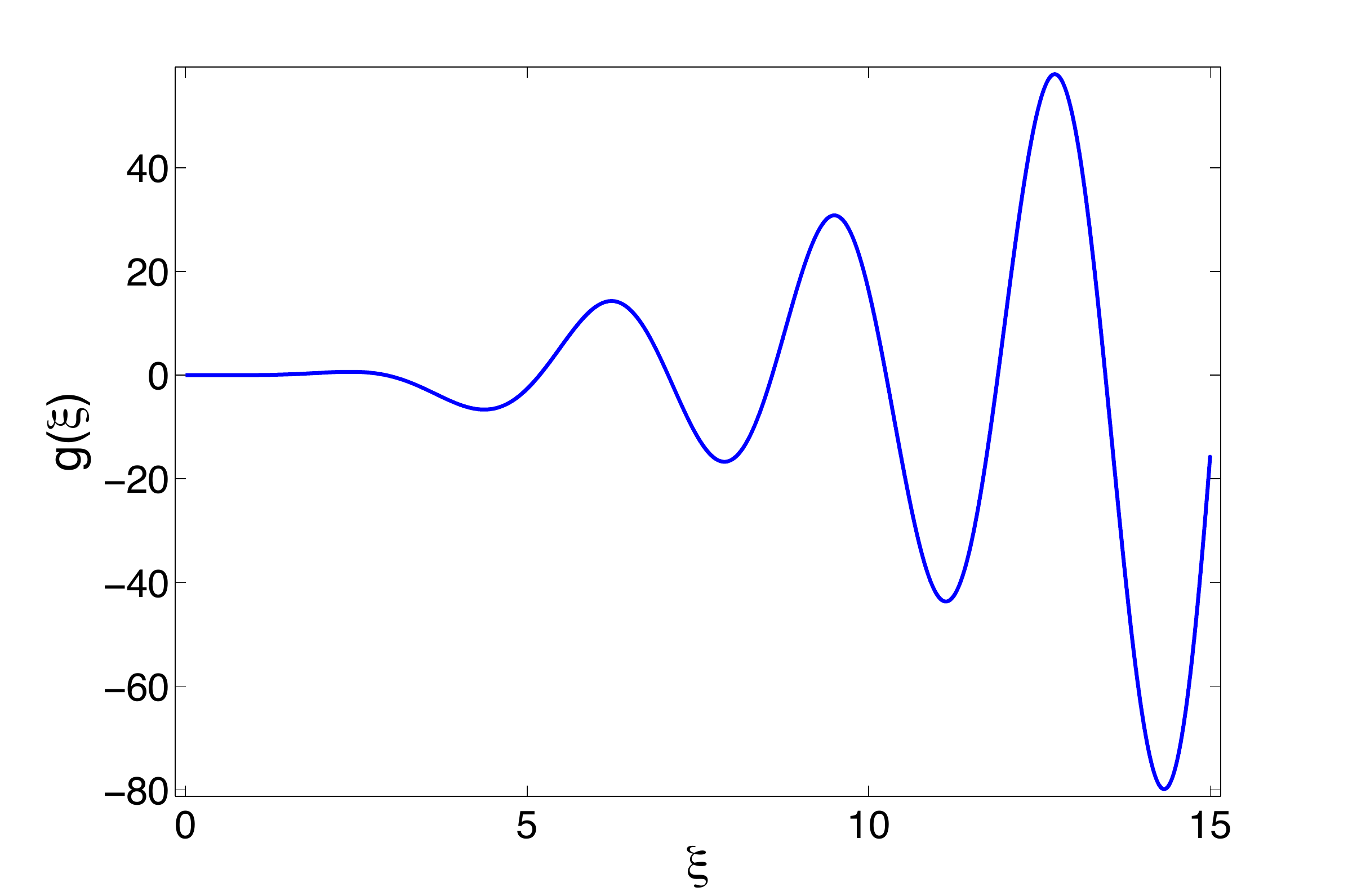}
\caption{(Color online) The function $g(\xi)$ ($\xi \equiv \frac{\omega a}{c}$),
defined for a given mechanical frequency $\omega$. Parameters 
of the autocorrelation function are  $k_{\rm min} = 0$ and $k_{\rm max} > \omega/c$.}
\label{fig2}
\end{figure}

We have discarded in $F_2$ an infinite (additive) contribution which is proportional
to $q(t)$. This divergence is due to the zero with of the mirrors, and can
naturally be interpreted as a renormalization of the self-energy of the
moving mirror. Besides, note that is not related to dissipation, since the
force is conservative, and can be derived from a harmonic potential.

In order to gain some insight into the nature of the system, we now write
$f^{(1)}$ in Fourier (frequency) representation,
\begin{equation}
\widetilde{f^{(1)}}(\omega)\;=\; \frac{4\pi}{a^2} \,
\tilde{q}(\omega)\; \int \frac{d^2{\mathbf
k}_\parallel}{(2\pi)^2}\, {\widetilde{\Omega}}({\mathbf k}_\parallel)
\,|{\mathbf k}_\parallel|^3\,\coth(|{\mathbf k}_\parallel| a)\,
( 1- a \frac{d}{da}) \,
\sum_{-\infty}^{+\infty} \frac{1}{(\frac{n \pi}{a})^2 + {\mathbf
k}_\parallel^2 - (\frac{\omega + i
\eta}{c})^2} \;. 
\end{equation}

Performing the summation, assuming rotational invariance for
$\widetilde{\Omega}$, and keeping just the dissipative terms, we see that
\begin{equation}
\widetilde{f^{(1)}}(\omega) \;=\; 2 \, \tilde{q}(\omega)\; \int_0^\omega
dk_\parallel  {\widetilde{\Omega}}(k_\parallel) \,
k_\parallel^4\,\coth(k_\parallel a) \,\csc^2(\sqrt{\omega^2- k_\parallel^2}
a) \;.
\end{equation} 
We write this expression in terms of a new function, 

\begin{equation}
\widetilde{f^{(1)}}(\omega) = \tilde{q}(\omega)\;  g(\xi) \; ,
\end{equation}
where, again, $\xi = \omega a/c$. In Fig. 2 we show the function $g(\xi)$
for the two-cutoff model autocorrelation function in which, 
for the sake of simplicity, we have set $k_{\rm min} = 0$ and $k_{\rm max}
> \omega/c$. Again, the existence of resonances is evident. As in the case
of the first example, we see that bigger frequencies are able to excite
all the normal frequencies ($\sqrt{(\frac{n \pi}{a})^2 + |{\mathbf
k}_\parallel|^2}$) which are smaller than $\omega$. That explains the fact
that the peaks grow with the frequency. 

Finally, as another manifestation of the same effect, we can derive the
spectral density of the power (to the second order
in $q$):
\begin{equation}
\widetilde{\dot{\mathcal E}^{(2)}}(\omega) \,=\, \int_0^\infty
\frac{d\omega}{2\pi}
\,{\mathcal P}(\omega)
\end{equation}
which has a rather similar interpretation to the one we calculated in the
previous example.  Its form can be obtained explicitly by performing the integration, the
result being:
\begin{equation}
{\mathcal P}(\omega) \;=\; \frac{8\pi}{a^2} c^2 \, |\tilde{q}(\omega)|^2 \, 
\int \frac{d^2{\mathbf
k}_\parallel}{(2\pi)^2}\, {\widetilde{\Omega}}({\mathbf k}_\parallel)
\,|{\mathbf k}_\parallel|^3\,\coth(|{\mathbf k}_\parallel| a)\,
( 1- a \frac{d}{da}) \,
\sum_{-\infty}^{+\infty} \delta\big[\omega - \sqrt{(\frac{n \pi}{a})^2 + |{\mathbf
k}_\parallel|^2}\big]\;. 
\end{equation}
The presence of the resonances is again evident, now in the 
$\delta$-functions.
The position of the resonances agrees with
the ones determined in Ref.~\cite{Kardar:1997cu} within the context of the DCE,
for a quantum system with a similar geometry but corresponding to a quantum
real scalar field. 
\section{Conclusions}\label{sec:conclu}
We have presented results about the spectrum of the radiation generated by
two closely related systems, containing a planar dipole layer distribution
which oscillates in front of a conducting plane, in terms of the layer
autocorrelation function. 
In one of the systems, where the layer on the moving plane is meant to
describe a dielectric with permanent polarization,  we have found that the
radiated power exhibits enhancements for certain frequencies of
oscillation. We interpret them as being related to the existence of
constructive interference between the radiation generated by the layer and
the one that bounces on the conductor before reaching a given point. 

In the other system, the  moving plane also has conductor boundary
conditions on the face opposite to the static plane, and is used to
describe patch potentials. The extra boundary condition makes it possible
to have infinite bounces, and therefore resonances, which we have plotted
for a simple autocorrelation function. The physical observables considered
for this case were the force exerted on the moving plane, and its power.
\section*{Acknowledgements}
We thank Prof.~F.~D.~Mazzitelli for many useful comments.
This work was supported by ANPCyT, CONICET, UBA, and UNCuyo.

\end{document}